\documentclass[letterpaper]{article}
\usepackage{amsmath}
\usepackage{algpseudocode}
\usepackage{algorithm}
\usepackage{graphicx}

\begin {document}
\title{Variance based Scheduling to Improve the QoS Performance at the Cell Edge }
\author{Trshant Bhat and Sanjay Singh\thanks {Sanjay Singh is with the Department of Information and Communication Technology, Manipal Institute of Technology, Manipal University, Manipal-576104, INDIA, E-mail: sanjay.singh@manipal.edu}}

\maketitle
  \begin{abstract}
Now a days mobile phones are most often used for data communication rather than voice calls. Due to this change in user behavior, there is a need to improve the QoS received by the user. One of the ways of improving the QoS is an efficient scheduling algorithm which incorporates the needs of the users and variation in channel condition. The parameters used to measure the efficiency of the scheduling algorithms are the Jain Fairness Index and the overall system throughput.
In this paper we have proposed a variance based scheduling algorithm which selects the user who has the highest variance of data transmitted in a given time frame as a parameter for scheduling. This ensures that eventually, the users transmit almost equal amounts of data regardless of channel condition.
The simulation results shows that the proposed algorithm achieves high Jain Fairness Index of 0.92 with a lesser drop in the system throughput 18\% as compared to Dynamically altering Proportionally Fair Algorithm's 20\% using the Proportionally Fair Algorithm as reference.
  \end{abstract}
  
 \section{Introduction}

Mobile phones were first introduced with the view of having usability of telephones but with the convenience of mobility. The infrastructure is primarily based on radio communication but multiple users are simultaneously supported by slotting them by frequency or time to improve the number of users supported. This method has limitations in transmission but since the human ear is resilient to slight errors in transmission of voice, this was considered enough originally. 

As time passed by and there was innovation in mobile Internet technologies (i.e., GPRS, 3G, 4G) accompanied by improvements in compression technologies (mp3, flv); the use of mobile phones expanded to much more than just telephony. However, data transmission now had to be error free, or in its absence, the user would have to transmit more data to replace the erroneously transmitted data. The infrastructure is split into cells for better management, but due to propagation losses and multipath fading, as the distance of the user from the cell increased, the capacity of the user to transmit data decreased. While the effect of this is negligible in voice communication, this negatively affected data transmission for users on the periphery of the cell. The advances in transmission technologies has decreased the transmission errors, but there is a need for improved scheduling algorithms that would satisfy all the users in all parts of the cell.
\par
Proportionally Fair Algorithm (PFA) \cite{p:pf} tries to solve this by using the concept of proportional fairness to schedule the users, but it eventually schedules all the users equally, causing the users in different parts of the cell to send varying amounts of data. Dynamically Updating Proportionally Fair Algorithm (DPFA) \cite{p:dpf} tried to increase the throughput of the users on the cell edge by restraining the users at the cell center, but this created unfairness to the users at cell center, along with a drop in the overall system throughput.

The primary objective of this paper is to demonstrate the use of variance in addressing the limitations of the scheduling algorithms as described above.

Rest of the paper is organized as follows. Section \ref{parameters} explains the evaluation parameters used for comparing the efficacy of a scheduler. Section \ref{algos} explains the existing scheduling algorithms PFA and DPFA. Section \ref{ia} describes the proposed algorithm. Section \ref{simu} discusses the simulation results and finally section \ref{conc} concludes the paper.

\section{Scheduling Algorithm  Evaluation Parameters}
\label{parameters}
Parameters used to evaluate the efficacy of scheduling algorithms are:
\begin{enumerate}
\item Overall System Throughput
\item Fairness Index
\end{enumerate}
The system throughputs is a counter maintaining account of the overall system throughput.
  \subsection*{Fairness Index}\label{fi}
  The concept of fairness \cite{p:fi} in mobile systems arises when a limited amount of bandwidth is to be shared among several users simultaneously. It is highly desirable to find a method and, preferably, an index to measure and compare the degree of fairness of a particular allocation policy. This index is called the Fairness Index (FI). The Fairness Index and throughput maximization are two main performance metrics that are often used to analyze and compare the performance of different resource allocation schemes, since these two metrics define the service provider and consumers satisfaction index respectively.

Jain fairness index \cite{jain} is given by,

\begin{equation}
	I_{Jain} = \frac{ \mid \sum_{i=1}^{n} \overline{r}_{i} \mid^{2} } {N\sum_{i=1}^{n} \overline{r}_{i}^{2} }
\end{equation}
where $\overline{r}_i$ is the throughput of the $i$th users.

The Jain Fairness index is used to measure the FI because of its computational simplicity. As per the definition of fairness index \cite{jain}, the system would be fairest when the throughput of the users would be equal.

\section{Existing Scheduling Algorithms}
\label{algos}
In this section, we describe and discuss two of the existing scheduling algorithms PFA and DPFA respectively.
 \subsection{Proportionally Fair Algorithm}\label{pfa}
In the Proportionally Fair Algorithm (PFA) \cite{p:pf}, the selection of the user at any time is based on a balance between the current possible rates and fairness. This algorithm performs scheduling by calculating the ratio of given rate for each user with their average throughput to date, and selecting the one with the maximum ratio, that is
\begin{equation}
P_{k(PFA)} = \arg\max_{k} \frac{r_{k}}{R_{k}}
\end{equation}
where $r_{k}$, in the numerator represents the instantaneous data rate of the $k$th users while $R_{k}$ represents the average rate of the same user based on the past resource allocation. The PFA allocates the resource at a given time to the user who has the maximum value of $P_{k(PFA)}$.
The average throughput $R_{k}$ is defined \cite{p:pf} as,
\begin{equation}
R_{k}[t] = 
	  \begin{cases}
	      ( 1 - \frac{1}{T_c} ) R_k[t-1] + \frac{1}{T_c} r_k[t],	& \text{if $k$ is scheduled} \\
	      ( 1 - \frac{1}{T_c} ) R_k[t-1],					& \text{if $k$ is not scheduled}
  \end{cases}
\end{equation}
where $T_c$ is the system time count which is initialized to 0 and increments every time slot.

The rate $R_{k}$ is updated every time slot after choosing the user. The rate updating equations are observed to update such that the average rate of the chosen user drops drastically while the unchosen user drops a little. As a result, in the next round of selection, the priority of the unchosen user increases exponentially whereas the priority of the chosen user increases marginally.

The PF scheduling is simple but effective for non real-time services. However it has been observed that this algorithm schedules all the users equally, hence the user at the cell edge would not be able to transfer as much data as the user at the cell center due to the degraded channel condition. This algorithm starves the users at the cell edge. Hence degradation in QoS performance. There is also a disparity in the individual user throughput.

\subsection{Dynamically Updating Proportionally Fair Algorithm}\label{dpfa}
Ning et al \cite{p:dpf} has proposed the modification to PFA. The drawbacks of the PFA are remedied by using two  controlling factors $\alpha$ and $\beta$ respectively, which is used as,
\begin{equation}
P_{k(DPFA)} = \arg\max_{k} \frac{ (r_{k})^\alpha }{ (R_{k})^\beta }
\end{equation}
where , \\
if $\alpha = \beta = 1$, then we get PF scheduling\\
if $\alpha = 1, \beta = 0$, then it is a maximum C/I scheduling \cite{p:maxci}\\
if $\alpha = 0, \beta = 1$, then it is a round-robin (RR) scheduling \cite{p:rr}

The value of $\alpha$ remain a small positive value usually 1, and the $\beta$ value will change for every user. There are two timers introduced in this algorithm, $A_{k}[t]$ and $B_{k}[t]$ to record the duration in which the user remain in the cell edge and at the cell center, respectively. $A_{k}[t]$ and $B_{k}[t]$ have been defined as,

\begin{equation}
A_{k}[t] =
	\begin{cases}
		0,              & \text{$\gamma_k[t] \geq \delta$}\\
		A_{k}[t-1] + 1, & \text{$\gamma_k[t] <    \delta$}
	\end{cases}
\end{equation}
and
\begin{equation}
B_{k}[t] =
	\begin{cases}
		0,              & \text{$\gamma_k[t] >    \delta$}\\
		B_{k}[t-1] + 1, & \text{$\gamma_k[t] \leq \delta$}
	\end{cases}
\end{equation}
where $\delta$ is the cell SNR threshold and $\gamma_k[t]$ is the SNR of the $k$th user.
Hence with these two equations, $\beta$ is updated as
\begin{equation}
\label{dpfa:beta}
\beta = 
    \begin{cases}
		1,              & \text{$A_{k}[t] \geq  \theta$ or $0 \leq B_{k}[t] \leq \theta$}\\
		\max \lbrace \frac{\gamma_k[t]}{\delta} , b \rbrace , 
		                & \text{$0 \leq A_{k}[t] \leq \theta$ or $B_{k}[t] >  \theta$}
	\end{cases}
\end{equation}
for all $k$ users. Parameter $\theta$ is the cell time threshold and $b$ is a constant usually equal to $0.5$. As per equation \ref{dpfa:beta}, the users in the cell edge are compensated and the users in the cell center are punished. Since the users at the cell edge are given higher priority in this algorithm, the throughput of those users is increased at the expense of the users at the center. Due to the fact that the users at the edge are scheduled more often, whose datarate is low as compared to users at cell center, the overall system throughput decreases as compared to the pure PF algorithm. The disparity in the individual user throughput reduces, increasing the FI to 0.86.

\section{Proposed Scheduling Algorithm}
\label{ia}
Following points has been considered while formulating the algorithm:
\begin{enumerate}
\item As the distance from the tower increases the user is capable of sending lesser data.
\item Punishing and compensating should be kept to a bare minimum.
\item It should be computationally simple.
\end{enumerate}
In rest of the paper, we will call the proposed algorithm as VPFA (Variance based Proportionally Fair Algorithm). We first define two counters,
\begin{enumerate}
\item \textbf{Individual User Throughput}: This counts the throughput of the individual user.
\item \textbf{FI Stability Counter ($C_s$)}: This counts the evaluations done for FI without any change in FI outside the required change.
\end{enumerate}
and two constants,
\begin{enumerate}
\item \textbf{Step interval between measurement of FI ($S_{fi}$)}: Number of time slots between which FI is evaluated.
\item \textbf{FI Stability Counter limit ($L_{sc}$)}: The count $C_s$ should reach for the algorithm to change from PFA to the improved algorithm.
\end{enumerate}
If we look at the evaluation of FI every few steps, we observe that it goes through a damping process before increasing in infinitesimally small amounts. Since the number of iterations required to improve to a good FI in PFA is large, we can change the selection criteria to ensure that data sent by each user remains the same. Since the the change in selection criteria should occur after the FI stabilizes, we use a FI stability counter $C_s$ which is triggered after the change in FI is lower than 0.01 and then iterate continuously till it reaches the limit set by the constant FI Stability Counter limit ($L_{sc}$). Once this limit is reached, the user is selected who has the highest throughput variance. Mathematically it can be written as

\begin{equation}
P_{k(VPFA)} = \arg\max_{k} \sigma^2_k
\end{equation}	
where $\sigma^2_k$ is the throughput variance of the $k^{th}$ user. The pseudocode of the proposed scheduling algorithm, VPFA is given in Algorithm \ref{alg:1}.
\begin{algorithm}[bpht!]
\caption{Improved Scheduling Algorithm (VPFA)}
\label{alg:1}
\begin{algorithmic}
\Procedure {Select\_Using\_PF}{}
\State Select user using pure PF
\If {$T_c~\%~S_{fi}~==~0~$}
    \State evaluate FI
    \If {$C_s == 0$}
    	\State $C_s = 1$
	\Else
    	\If {$LastFI - FI < 0.01$}
        	\State increment $C_s$
        \Else
        	\State $C_s = 0$
    	\EndIf
    \EndIf
    \State $LastFI = FI$
\EndIf
\If {$C_s~==~L_{sc}$}
	\State $GOTO$ Select\_Using\_LDS
\Else
	\State $GOTO$ Select\_Using\_PF
\EndIf
\EndProcedure
~\\
\Procedure {Select\_Using\_Variance}{}
	\State Select user with highest variance
	\State $GOTO$ Select\_Using\_Variance
\EndProcedure

\end{algorithmic}
\end{algorithm}

\newpage
\section{Simulation Results and Discussion}
\label{simu}
The simulation of the existing and proposed scheduling algorithms has been programmed using Python 2.7 \cite{py} using the NumPy \cite{npy} and SciPy \cite{spy} libraries. The simulation parameters are given in Table \ref{tab:simpar}.

\begin{table}[bpht!]
		\centering
		\caption{Simulation Parameters}
		\label{tab:simpar}
		\begin{tabular}{|l|l|}
		\hline
			\textbf{Parameter} & \textbf{Value} \\ \hline
			Distance-dependent path loss & COST-231 \\ \hline
			Shadowing fading & lognormal distribution \\ \hline
			User State & Static Positions \\ \hline
			Shadowing standard deviation & 8 dB \\ \hline
			Total downlink TX power in traffic channels & 46 dBm \\ \hline
			Cell radius & 1km \\ \hline
			System carrier frequency & 2GHz \\ \hline
			System bandwidth & 10MHz \\ \hline
		\end{tabular}
\end{table}

\begin{figure}[bpht!]
    \includegraphics[height=8cm,width=12cm]{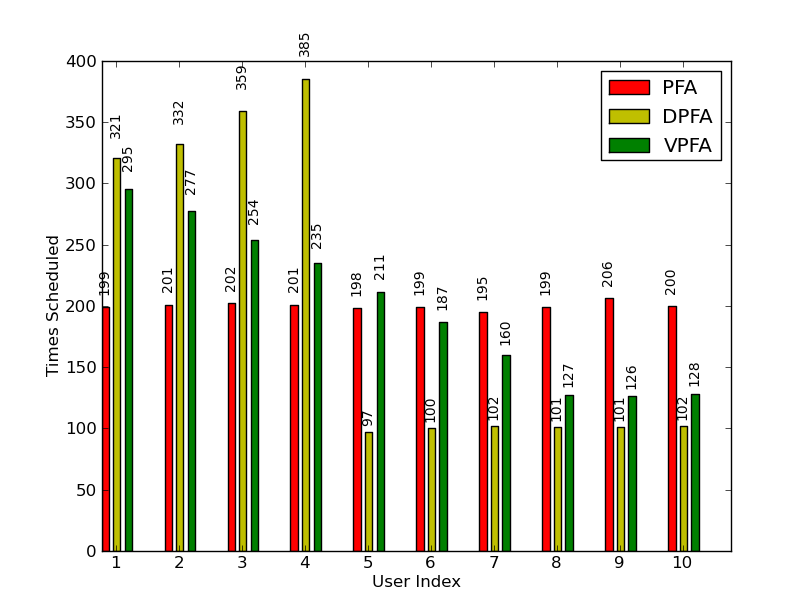}
    \caption{Comparisons of number of times individual users have been scheduled for all PF algorithms}
    \label{cri:g:ics}
\end{figure}

\begin{figure}[bpht!]
    \includegraphics[height=8cm,width=12cm]{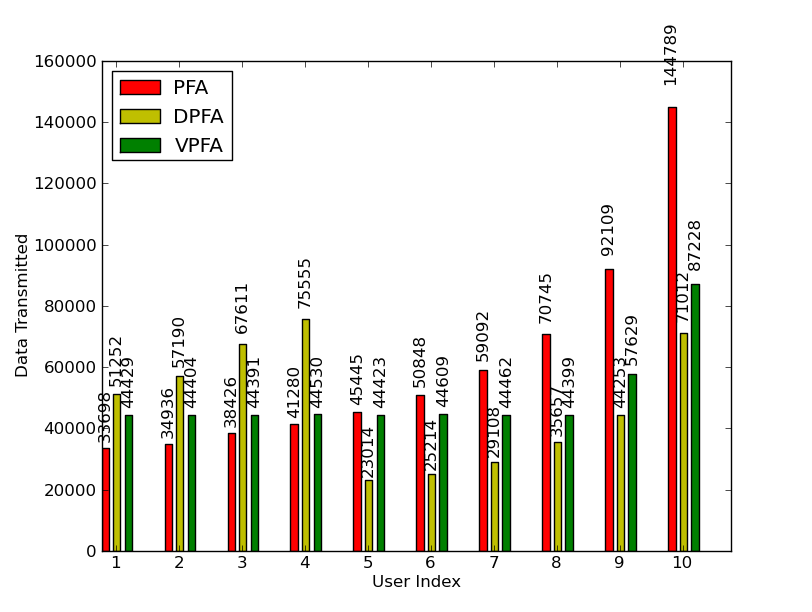}
    \caption{Comparisons of throughput of individual users for all PF algorithms}
    \label{cri:g:it}
\end{figure}
\begin{figure}[bpht!]
    \includegraphics[height=8cm,width=12cm]{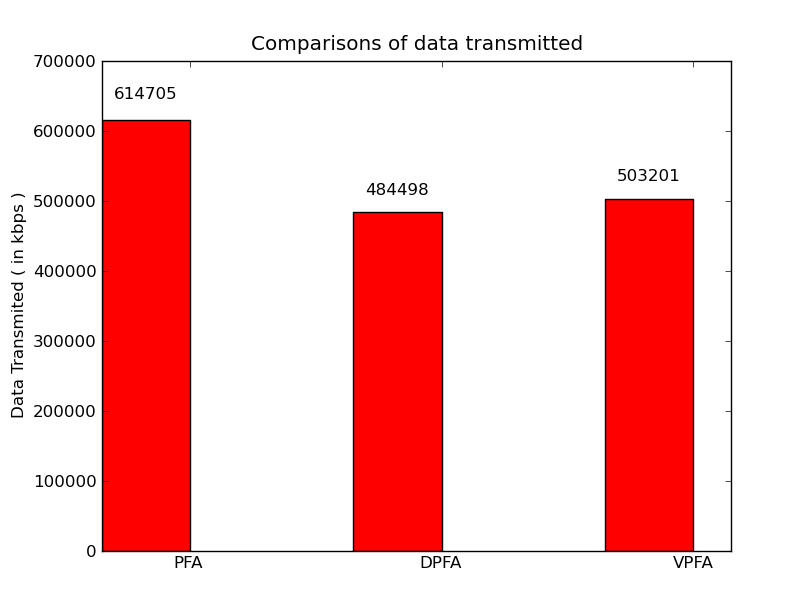}
    \caption{Comparisons of system throughput for all PF algorithms}
    \label{cri:g:st}
\end{figure}

\begin{figure}[bpht!]
    \includegraphics[height=8cm,width=12cm]{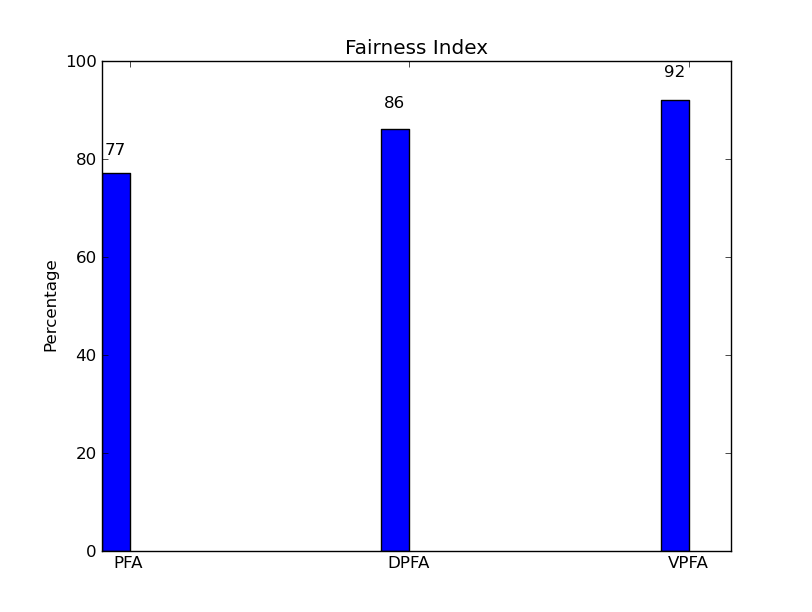}
    \caption{Comparisons of FI for all PF algorithms}
    \label{cri:g:fi}
\end{figure}

The simulation parameters were kept constant in simulating all the algorithms so as to provide the most fair evaluation of all the algorithms. The user index increases with the decrease in distance from the tower. 
Figure \ref{cri:g:ics} shows how the users were scheduled for all the algorithms. It is observed that the users are scheduled almost equally in the PFA. DPFA schedules the cell edge users well, but punishes the cell center users for it. The proposed algorithm schedules the user most away from the tower most of the times and the number of times the user was scheduled drops as the distance from the center reduces. This causes the users to transmit almost the same amount of data is shown in Fig.\ref{cri:g:it}.
\par
Considering the effect on the overall system, it is observed that even though the drop for the improved algorithm is 18\% considering the PFA as reference, it is still better than the 20\% drop in overall system throughput of the DPFA. Due to the fact that all the users transmit almost the same amount of data, the Fairness Index of the proposed algorithm is 0.92 which is better than the FI obtained by DPFA which is 0.86 also there is substantial improvement over PFA's 0.77. The throughput of the user at the cell edge improves by 30\%. 
However, when the variance based selection criteria is working, the user with the smallest variance is starved. This is a drawback of the algorithm.
From Fig.\ref{cri:g:st}, we see that the drop in overall system throughput for the VPFA is 18\% as compared to DPFA's 20\%. PFA is used as reference.
Figure \ref{cri:g:fi} shows the FI achieved by various algorithms.

\section{Conclusion}\label{conc}
In this paper by using variance of the individual user data transmitted as the selection factor in scheduling the users we are able to equalize the data transmitted by every user in the cell regardless of distance from the tower. This causes a high Fairness Index of 0.92 which is significantly higher than those obtained through PFA and DPFA. There is a appreciable improvement in the throughput of the users throughout the cell. By adopting the proposed scheduling algorithm, the QoS offered by the telecommunication companies can be improved which may lead to reduction in user attrition to other telecoms service provider hence gain in company revenue. However, the matter of starvation of the user who has the maximum throughput in the cell has scope for future research. 


\bibliographystyle{IEEEtran}
\bibliography{ref}

\end{document}